**An improved model of metal/silicate differentiation during Earth's accretion**

**Authors**: *K.I. Dale[1], D.C. Rubie[2], M. Nakajima[3], S. Jacobson[4], G. Nathan[4], G.J. Golabek[2], S. Cambioni[5], A. Morbidelli[1]*

[1]Laboratoire Lagrange, CNRS, Observatoire de la Côte d'Azur, Université Côte d'Azur, Nice, France
[2]Bayerisches Geoinstitut, University of Bayreuth, 95490 Bayreuth, Germany
[3]Department of Earth and Environmental Sciences, University of Rochester, Rochester, USA
[4]Department of Earth & Environmental Sciences, Michigan State University, East Lansing, MI 48824, USA
[5]Department of Earth, Atmospheric and Planetary Sciences, Massachusetts Institute of Technology, Cambridge, MA, USA

**Abstract:** We improved the algorithm presented in Rubie et al. (2015) to model the chemical evolution of Earth driven by iron/silicate differentiation during the planet's accretion. The pressure at which the equilibration occurs during a giant impact is no longer a free parameter but is determined by the smooth particle hydrodynamic (SPH) simulations of Nakajima et al. (2021). Moreover, impacting planetesimals are now assumed to be too small to cause melting and differentiation and thus their materials are stored in the crystalline upper mantle of the growing planet until a hydrostatically relaxed global magma ocean forms in the aftermath of a giant impact, whose depth is also estimated from Nakajima et al. (2021). With these changes, not all dynamical simulations lead to a satisfactory reproduction of the chemical composition of the bulk silicate Earth (BSE). Thus, the latter becomes diagnostic of the success of dynamical models. In the successful cases also the BSE abundances of W and Mo can be reproduced, that were previously hard to fit (Jennings et al., 2021).

1. **Introduction.**

Rubie et al. (2015) coupled, for the first time, the geochemical evolution of the growing Earth to its accretion history as modeled in dynamical simulations. The dynamical simulations provide the genealogical tree of a planet, recorded as the history of impacts of planetesimals and planetary embryos that delivered the planet's mass. In the Rubie et al. (2015) code the initial chemical compositions of planetesimals and embryos are linked to their original birthplaces via a disk compositional model. Both planetesimals and embryos are assumed to be differentiated (iron-rich core surrounded by a silicate-rich mantle). Embryos also evolve due to their own collisional history before their final impact on the planet. During an impact, the core of the impactor chemically equilibrates with part of the mantle of the target. The fraction of the target's mantle involved in the equilibration process is determined by the hydrodynamic model of Deguen et al. (2014). The pressure at which equilibration occurs is assumed to be a fraction of the pressure at the core-mantle boundary of the target at the time of impact. That fraction is a free parameter that is determined by optimizing the final fit of the results to the bulk silicate Earth (BSE) composition. The pressure is associated with a temperature which lies approximately midway between the solidus and liquidus temperatures of peridotite at that pressure (Rubie et al., 2015). Once pressure and temperature are determined, partition coefficients extrapolated from laboratory experiments combined with a mass balance approach set the fractions of siderophile elements of the reacting materials that combine with the iron metal of the projectile's core and are transported into the target's core. Part of the iron of the projectile's core can also be oxidized by the reaction with the silicate and water and remains as FeO in the target's mantle. For all equations and details, we refer to Rubie et al. (2011, 2015).

Rubie et al. (2015) demonstrated their approach on 6 dynamical simulations of Earth formation in the so called "Grand Tack" scenario (Walsh et al., 2011; Jacobson and Morbidelli, 2014). They showed that with appropriate assumptions on the radial distribution of chemical properties of the disk and equilibration pressure during impacts, the BSE chemical composition in terms of FeO, $SiO_2$, Ni, Co, Nb, Ta, V and Cr could be satisfactorily reproduced in all simulations. In essence, the number of free parameters helps find a good fit irrespective of the history of the planet. In this situation, the chemistry of Earth cannot be used to discriminate between different dynamical models.

In this Note we propose two improvements to the original code of Rubie et al. that eliminate the use of a crucial free parameter: the pressure of metal/silicate equilibration. The first improvement is that, when a giant impact occurs, as in Gu et al. (2023), we now compute the depth (and hence the basal pressure) of the magma pond produced by the impact, using the results of the SPH impact simulations of Nakajima et al. (2021). A second improvement concerns the modeling of the equilibration of planetesimals' material. Each planetesimal in the dynamical simulations represents a swarm of smaller objects that are generally too small to induce significant melting of the target. Thus, regardless of whether they are differentiated or not, their material is stranded in the crust and the upper portion of the target's mantle until a giant impact occurs. This assumption is also made in Gu et al. (2023). Differently from that paper, however, we do not assume that planetesimal material is equilibrated at the base of the magma pond induced by the giant impact. Because of its longitudinal and latitudinal dispersion, we assume that planetesimal material can equilibrate only after the magma pond hydrostatically relaxes into a global magma ocean (MO). At this point, the metal previously delivered by planetesimals is dispersed as small droplets and progressively segregates, with equilibration occurring in a boundary layer at the base of the MO whose depth and basal pressure are also estimated from Nakajima et al. results. We also modify the calculation of the volume of the target silicate equilibrating with the impacting embryo's metal using the new experimental results of Landeau et al. (2021), whereas Gu et al. (2023) assumed a fixed fraction, independent on the collision conditions.

After detailing these changes in Sect. 2, we show in Sect. 3 that not all the 6 dynamical simulations considered in Rubie et al. lead to "good BSEs". The removal of equilibration pressure as a free-parameter limits significantly the capability of reproducing the BSE composition regardless of the accretion history. The results now depend crucially on the sequence of impacts and their associated energies. Thus, the chemical properties of Earth become diagnostic of the success of distinct dynamical simulations. We expect that this is not only true for the Grand Tack model, but also for other dynamical models, and maybe will even allow rejecting some models as never being successful. This will be the object of future work. In addition, we show that some of the Grand Tack simulations also reproduce the BSE concentrations of W and Mo without the need to assume an unconventional sulfur-poor and carbon-enriched Earth composition, as suggested by Jennings et al. (2021). This important result is the merit of the more realistic treatment of planetesimal impacts.

2. **Technical description of the improvements.**

Nakajima et al. (2021) used more than 100 SPH simulations to calibrate a scaling law predicting the depth of the magma pond produced in giant impacts involving projectile/target mass ratios in the range 0.03-0.5 and total masses (target + projectile) in the range 0.1-5.3 Earth masses, for a discrete set of impact angles (0°, 30°, 60°, and 90°, the latter being a grazing impact). For the purpose of this work, those results have been extended to predict also the depth and mass-fraction (relative to the target's mantle) of the global MO that is produced after hydrostatic relaxation and by computing the scaling law also for an impact angle of 45°. The python script implementing the scaling law, available at github.com/mikinakajima/MeltScalingLaw, has been converted into a C++ module (available on demand) that is called within the metal-silicate equilibration code of Rubie et al. (2015) as the latter works through the sequence of giant impacts. All the giant impacts recorded in the simulations of Rubie et al. (2015) lie within the total masses and mass ratios covered by the scaling law. Here, each impact angle in the simulations is rounded to the nearest value for which the scaling law is available. Also, the scaling law is defined for two values of the entropy of the target, that are 3160 J/K/kg (corresponding to a surface temperature of 2000 K, just below the melting temperature - hot target), and 1100 J/K/kg (corresponding to a surface temperature of 300 K - cold target). In our code we decide at input if all impacts occur on hot or cold targets, given that the cooling of the embryos in between giant impacts is not modeled. The depth of the magma pond resulting from each giant impact is then converted into a pressure and a temperature at the base of the pond, which are used in the equations that determine equilibration (see Rubie et al., 2011, 2015) between the fraction of the target's pond given by Deguen et al. (2014) and the impactor's core.

Assuming that the target is solid (giant impacts producing magma oceans are statistically spaced in time much more than the typical ~Myr MO crystallization times estimated by Elkins-Tanton,2008), planetesimals are unlikely to cause melting on a planetary embryo regardless of their velocity and impact angle because of their low mass relative to the target's mass. Thus, no metal-silicate reaction is implemented immediately after a planetesimal impact, but the chemical information on the delivered material, assumed to be statistically uniformly distributed in the near-surface of the target, is added to a cumulative vector that represents the contributions from all the planetesimals that hit a given embryo between giant impacts. When a giant impact occurs, after the equilibration of projectile's core with part of the target's mantle as described above, we assume that the silicate melt relaxes hydrostatically into a global MO, whose mass and depth are given by the Nakajima et al. scaling law. We assume that this MO contains (i) the material previously delivered by planetesimals to both colliding embryos and (ii) the fraction of the target's magma pond that equilibrated with the impactor's core. The remaining MO material comprises a uniform mixture of the unequilibrated portion of the target's mantle and the impacting embryo's mantle, which also constitutes the mantle's complement of the new global magma ocean. All the metal in the MO now comes from the planetesimal-delivered material.

Following the *fractionation model 2* of Rubie et al. (2003, p.253), previously unequilibrated metal delivered by planetesimals is assumed to become dispersed as small droplets in the MO. Final metal-silicate equilibration is assumed to occur in a mechanical boundary layer at the base of the magma ocean in which vertical convection velocities are close to zero (Martin and Nokes, 1988), at its characteristic pressure and temperature. To facilitate numerical calculations the boundary layer is assumed to comprise 5% of the magma ocean's total volume. The metal in that layer, after equilibrating with the silicate, is transferred to the target's core, while the equilibrated silicate in the layer and the unequilibrated silicate in the rest of the MO are mixed, assuming vigorous convection. This procedure is then iterated until no metal remains in the MO. The resulting silicate in the MO is then assumed to crystallize and mix with the rest of the mantle due to solid-state convection. Finally, the cumulative vector representing the material added by planetesimals to the target is reset, so that it will be ready to accumulate the planetesimal impacts delivered before the next giant impact. When a simulation is finished, if the cumulative vector is not empty it is identified with the late veneer and its content is simply added to the final mantle of the planet, with the metal being oxidized by water if enough water is available.

In Rubie et al. (2015), following Deguen et al. (2014), the volume of the target silicate that reacts with the projectile's metal was computed as that of a cone of radius $r(z)=r_0+\alpha z$, where $r_0$ is the radius of the projectile's core, z is the depth of the melt pool or magma ocean (that can range from 0 at the surface to the thickness of the mantle) and $\alpha$ is a coefficient with a value ~0.25. From a new series of laboratory experiments accounting for the jet launched during the impact, Landeau et al. (2021) derived a new formulation for $r_0$:

$$r_0 = R\, c_1\, [(\rho - \rho_s)/\rho_s]^{c_2}\, (U/U_l)^{2c_3}\, [2(1+R_t/R)]^{c_3}$$

where $c_1$, $c_2$ and $c_3$ are coefficients, U is the impact velocity and $U_l$ is the escape velocity from the combined projectile-target body of radius $R_t$, $\rho_s$ is the density of the silicate mantle. For a differentiated projectile, there is some ambiguity on what R and $\rho$ should be. If the Rayleigh instability fully mixes the core and the mantle of the impactor during the impact process, the impactor practically undifferentiates and therefore R should be the radius of the impactor and ρ its mean density. This assumption gives the maximal estimate for $r_0$. If instead the core-mantle mixing in the projectile is negligible, R should be identified with the projectile core radius and ρ should be the density of metal in the core. This gives the minimal estimate for $r_0$. The ratio between these two estimates is about 2. Reality probably lies between these two bounds, with the additional constraints that $r_0$ has to be larger than the projectile core radius and the volume of the cone of radius r(z) does not exceed the volume of the magma pond. In absence of laboratory experiments using bi-fluid projectiles, we consider both estimates in the following, comparing the results with those achieved with the original Deguen et al. (2014) recipe.

**Results**

We tested the updated versions of the model on the six N-body simulations studied in Rubie et al. (2015), for which we use the same name conventions. For each, we tested both assumptions of hot and cold targets. Then, for each of these two cases, we changed the calculation of $r_0$ from Deguen et al. (2014) to that of Landeau et al. (2021), considering both the maximal and minimal estimates discussed above.

The top panel of Fig. 1 shows how well the BSE is reproduced by the final calculated mantle composition of the model Earth in each simulation. We note that Gu et al. (2023) never attempted a global fit of the Earth's mantle composition. Here, as a first step, only the concentrations of FeO, $SiO_2$, Ni, Co, Nb, Ta, V and Cr, as well as the Nb/Ta ratio, are considered in the $\chi^2$ calculation, in order to make a direct comparison with the results of Rubie et al. (2015). We see that, while the original code always produces small $\chi^2$ values thanks to equilibration pressure being a free parameter, the updated code with the Deguen et al. (2014) recipe for the volume of reacting silicate (where the only remaining free parameters are those describing the distribution of metallic Fe and Si in the starting bodies as a function of their original heliocentric distance) finds equivalent $\chi^2$ values (within error bars) only in three hot-target simulations and two cold-target simulations across four of the six N-body simulations. Two N-body simulations, 4:1-0.25-7 and i-4:1-0.8-4 (see Rubie et al., 2015), fail to ever find a good fit using the updated version of the model, thus potentially eliminating them as valid simulations for the formation of Earth. For these simulations the best-fit equilibration pressure was much higher than that of the average magma ponds expected from the recorded giant impacts. The use of the Landeau et al. (2021) recipe for the volume of the reacting slicate gives almost identical results to those using the Deguen et al. recipe, if $r_0$ is computed assuming that R and ρ are the projectile core's radius and density. Instead, if we consider the full projectile with averaged density, the results are typically degraded, although we still find comparably good fits in one hot-target simulation and two cold-target simulations.

Jennings et al. (2021) extended the analysis to moderately siderophile Mo and W, finding marginally good fits for the former and bad results for the latter, unless an unconventional carbon-rich Earth was assumed. In our code, the C-concentration of the building blocks of Earth is not arbitrary: we follow Hirschmann et al. (2021) and Blanchard et al. (2022) in assuming that embryos have 0.004 wt % C in bulk, whereas undifferentiated non-carbonaceous and carbonaceous planetesimals[1] contain respectively 0.16 and 3.35 wt% C.  We assume that all planetesimals are undifferentiated, but assuming that 50% of non-carbonaceous planetesimals are differentiated and have the same carbon concentration as embryos does not change substantially the results. Our final Earths are consistent with the low C abundance BSE estimated in Hirschmann & Dasgupta (2009), Halliday (2013), Hirschmann (2018) and Blanchard et al. (2022).

Despite this low C content, the new version of the model is far more successful in its approach than previous attempts. For all simulations that produce a 'good Earth' in the top panel of Fig. 1 there is an improved fit for both Mo (middle panel) and W (bottom panel) compared to the original model. The new, more realistic treatment of metal-silicate fractionation in connection with planetesimal delivery is the reason for this improved result. Planetesimals feed dispersed metal droplets to the global MO, allowing a more substantial segregation of siderophile elements into the core than in the case where metal-silicate equilibration only occurs in a small fraction of the magma pond and at higher pressure, as in Jennings et al. (2021).

---

[1] In the Grand Tack simulations the carbonaceous chondritic planetesimals are defined as those initially located beyond Jupiter.

Rubie et al. (2016, Fig. 1) showed that metal-silicate segregation alone results in highly siderophile element (HSE) concentrations that are greatly in excess of BSE abundances and thus demonstrated the need for FeS exsolution and segregation (the hadean matte). We have included HSEs in our model to determine if the modifications presented here substantially reduce calculated HSE concentrations and thus eliminate the need for the hadean matte. The final calculated mantle abundances of HSEs are indeed lower in many simulations when using the new code in comparison to the results of the original code but are generally still much higher than BSE concentrations. These overabundances stress the importance of FeS exsolution (which is not included in the results presented here) as argued by Rubie et al. (2016).

3. **Conclusions**

We have updated the Rubie et al. (2015) core formation model for a growing planet. It now incorporates the results of Nakajima et al. (2021) using the information from N-body simulations to calculate the amount of melting in the mantle and the pressure at which material equilibrates after each giant impact. We also accumulate the materials delivered by planetesimals at the top of the target's solid mantle until a subsequent giant impact occurs, following which they equilibrate in a hydrostatically-relaxed global magma ocean. These changes allow us to remove the equilibration pressure as a free parameter of the code. Given the importance of pressure for the equilibration reactions this is an important change. Using the same disk composition free parameters as in Rubie et al. (2015), not all simulations result anymore in reproducing satisfactorily the chemical composition of the bulk silicate Earth, but they become sensitive to the energy and sequence of the giant impacts that built our planet. The chemistry of the BSE thus becomes diagnostic of the validity of a given dynamical scenario of Earth's accretion. In the successful simulations the new treatment of planetesimal accretion allows improving the match with the BSE abundances of Mo and W, which had been previously difficult to fit and reinforces the need for a hadean matte to explain the mantle HSE depletion.

**Acknowledgments**
This work has been supported by the ERC project Holyearth (N. 101019380). S.C. was supported by the Crosby Postdoctoral fellowship of the Department of Earth, Atmospheric and Planetary Sciences, Massachusetts Institute of Technology. M.N. was supported in part by the National Aeronautics and Space Administration (NASA) grant numbers 80NSSC19K0514 and 80NSSC21K1184. Partial funding for M.N. was also provided by NSF EAR-2237730 as well as the Center for Matter at Atomic Pressures (CMAP), an NSF Physics Frontier Center, under Award PHY-2020249. Any opinions, findings, conclusions or recommendations expressed in this material are those of the authors and do not necessarily reflect those of the National Science Foundation. M.N. was also supported in part by the Alfred P. Sloan Foundation under grant G202114194.

**Caption of Figure 1:**

Top: the value of $\chi^2$ of the fit of the simulation results to the BSE composition in terms of FeO, $SiO_2$, Ni, Co, Nb, Ta, V, Cr and Nb/Ta ratio. Error bars are 3 σ on the values of $\chi^2$, where $\sigma=(2v)^{1/2}/v$ and v is the number of elements fitted minus the number of fitting parameters. Six Grand Tack simulations are considered. The blue dots refer to fits obtained with the original Rubie et al. (2015) code where the equilibration pressure is a free parameter of the fit. Green triangles depict the fits obtained with the modified code, with planetesimal accumulation and equilibration pressure calculated using the results of Nakajima et al. (2021). Points in orange show the $\chi^2$ obtained considering the full projectile with averaged density and using the recipe of Landeau et al. (2021), while points in purple show the results using this recipe but considering only the radius of the projectile core and its metallic density. Darker points in all models represent simulations with a hot target and lighter points represent models where the target was cold. Values outside the range of the scale are indicated with arrows in their respective model colors and are labelled with the corresponding $\chi^2$.

Middle: The BSE abundance of Mo in all the simulations that give $\chi^2$ values in the top panel comparable, within error bars, with those of the original model of Rubie et al. (2015). Simulation i-4to1-0.8-4 is excluded as no version of the updated model results in a good fit. Colour scheme is as in the top panel. The error bars are derived by propagating the uncertainties of the experimentally derived partition coefficients. The actual BSE composition and its uncertainty are depicted by the horizontal line and grey band. The original model results (blue dots) and errors have been obtained using the version of the code used in Jennings et al. (2021) but assuming C content in the original embryos and planetesimals given by Hirschmann et al. (2021).

Bottom: the same as the middle panel, but for the W abundance in the BSE.

In most simulations, the inclusion of Mo and W in the fits increases moderately the value of reduced $\chi^2$, for instance from 1.7 to 4 in the case of simulation 4to1-0.5-8_HOT. But in some, for instance 4to1-0.25-7_HOT when the whole impactor is used in the Landeau et al. (2021) recipe, $\chi^2$ improves.

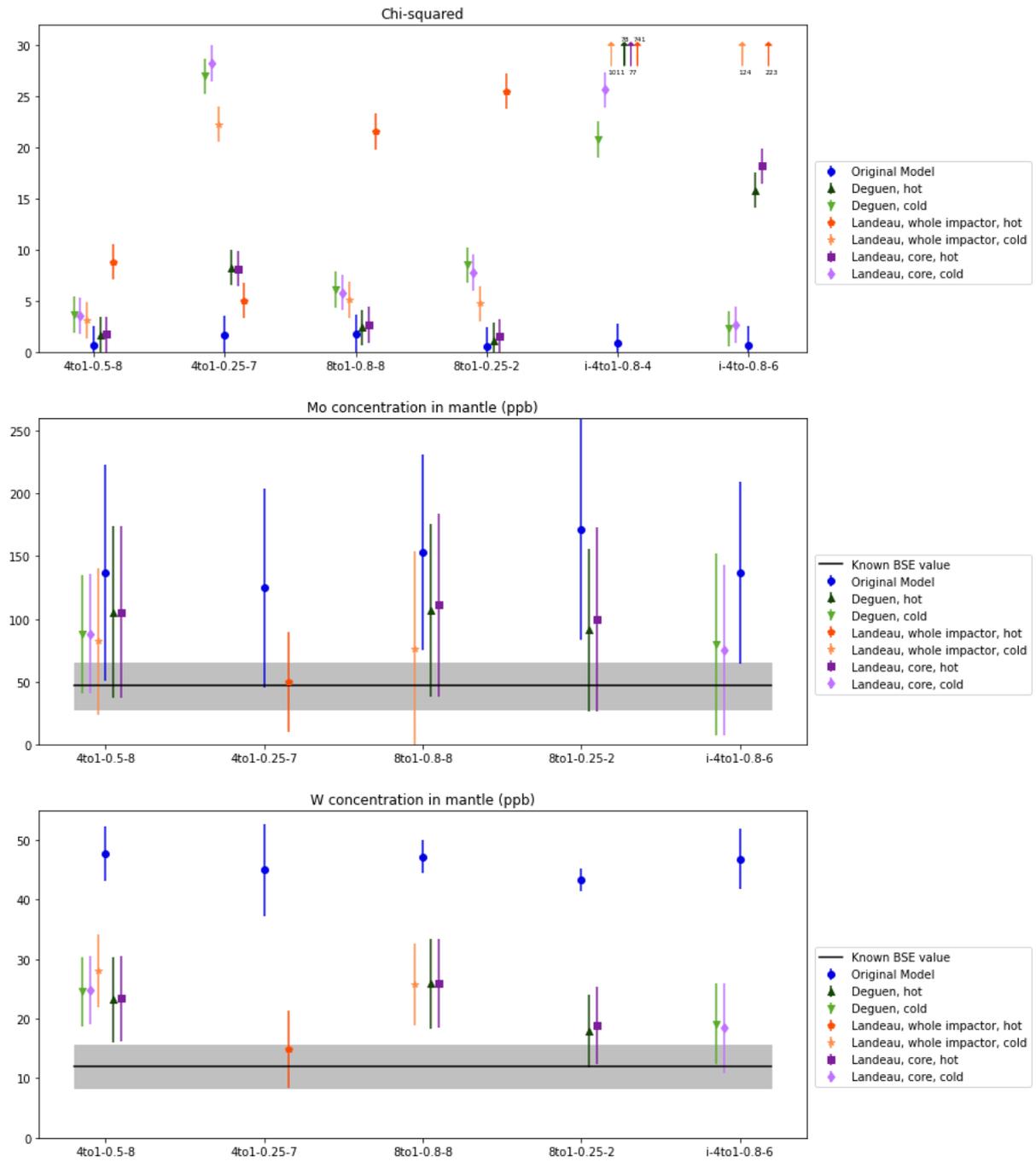

**Fig. 1**